\documentstyle[12pt,aasms4]{article}
\begin{document}

\title {GLOBAL DYNAMICS OF ADVECTION-DOMINATED ACCRETION REVISITED}
\author{Ju-Fu Lu$^{1,2}$, Wei-Min Gu$^1$, Feng Yuan$^3$}
\altaffiltext{1}{Center for Astrophysics, University of Science and
     Technology of China, Hefei, Anhui 230026, P. R. China}
\altaffiltext{2}{National Astronomical Observatories, Chinese Academy
     of Sciences}
\altaffiltext{3}{Department of Astronomy, Nanjing University, Nanjing, 
Jiangsu 210093, P. R. China}

\authoremail{lujf@ustc.edu.cn}

\begin{abstract}
We numerically solve the set of dynamical equations describing
advection-dominated accretion flows (ADAF) around black holes, using a
method similar to that of Chakrabarti (1996a). We choose the sonic radius of
the flow $R_s$ and the integration constant in angular momentum equation $j$
as free parameters, and integrate the equations from the sonic point inwards
to see if the solution can extend supersonically to the black hole horizon,
and outwards to see if and where an acceptable outer boundary of the flow
can be found. We recover the ADAF-thin disk solution constructed in Narayan,
Kato, \& Honma (1997, NKH97), an representative paper of the previous works
on global ADAF solutions, although in that paper an apparently very
different procedure was adopted. The use of our method has the following
advantages. First, we obtain all the solutions belonging to the ADAF-thin
disk class, not only some examples as in NKH97. Second, we find other
classes of solutions that were not noticed in NKH97, namely the ADAF-thick
disk solution in which an ADAF connects outwards to a thick disk, and the $ 
\alpha $-type solution which can extend either only to the black hole
horizon or only to the outer boundary. The ADAF-thick disk solution may have
astrophysical implications, in view that in some cases models based on the
ADAF-thin disk solution encounter some difficulties. The $\alpha $-type
solution is also worth studying, in the sense that such a solution could be
a part of a shock-included global solution. Apart from all these classes of
solutions, there are definite ranges of incorrect values of $R_s$ and $j$ for
which no solutions exist at all. Taking all these results together, we
obtain a complete picture in the form of $R_s-j$ parameter space which sums
up the situation of ADAF solution at a glance. For comparison we also
present the distribution of global solutions for inviscid flows in the $R_s
-j$ space, which supports the view that there should be some similarities
between the dynamical behavior of ADAF and that of adiabatic flows, and that
there should be a continuous change from the properties of viscous flows to
those of inviscid ones.
\end{abstract}


\section {Introduction}

Accretion of rotating matter onto a compact object powers many energetic
astrophysical systems such as cataclysmic variables, X-ray binaries, and
active galactic nuclei. In order to explain the observational radiative
output of these systems, a full understanding of the global structure and
dynamical behavior of accretion flows is of fundamental importance. Here the
word 'global' means the entire region of the flow, i.e. from the place where
accretion starts till the surface of the compact object. The detailed
microphysics of radiative cooling processes, although necessary for
calculating the emission characteristics, may or may not significantly
affect the flow dynamics, depending on whether the cooling is efficient or
not.

The birth of modern accretion disk theory is traditionally attributed to the
model of Shakura \& Sunyaev (1973). In this model the accretion flow was
considered to be geometrically thin (equivalently, thermally cool, cf. eq.
[3] below), optically thick, Keplerian rotating, and radially highly
subsonic. Such a simple model has been very successful, particularly for the
case of unmagnetized white dwarf accretion, i.e. in cataclysmic variables
(see Frank, King, \& Raine 1992 for a review). The reason for this fact is
that the gravitational potential well of a white dwarf is not very deep,
thus the radial velocity of the accreted matter is small, and the
temperature is low (both the kinetic energy and the thermal energy are
converted from the gravitational energy); such a physical situation just
meets the basic assumptions of Shakura \& Sunyaev model, and the model in
turn provides an excellent description of the global structure and dynamics
of the white dwarf accretion disk.

However, black hole accretion must be transonic, i.e. the radial velocity
must be important in the region near the black hole, and some or even almost
all the energy and entropy of the accreted matter can be advected into the
hole, rather than being radiated away from the surface of the flow (neutron
star accretion could be either transonic or entirely subsonic). Such a
transonic or advective feature cannot be reflected by Shakura \& Sunyaev
model at all, and adequate models should be developed. There are two known
regimes in which the radiative cooling is very inefficient and most of the
dissipated energy is advected into the black hole, namely, when the mass
accretion rate is very high and the flow is optically thick, and when the
accretion rate is very low and the flow is optically thin. The resulting
so-called advection- dominated accretion flows (ADAF) have been a very
attractive subject for the recent years, especially because of promising
applications of the optically thin ADAF model to explaining properties of
X-ray transients, the Galactic center, low luminosity active galactic
nuclei, and some other high energy objects (see Narayan, Mahadevan, \&
Quataert 1998 for a review). In particular, a number of papers have been
devoted to the study of global solutions of ADAF (Matsumoto, Kato, \& Fukue
1985; Abramowicz et al. 1988; Honma, Matsumoto, \& Kato 1991; Chen \& Taam
1993; Abramowicz et al. 1996; Chakrabarti 1996a; Igumenshchev, Chen, \&
Abramowicz 1996; Chen, Abramowicz, \& Lasota 1997; Nakamura et al. 1997;
Narayan, Kato, \& Honma 1997; Peitz \& Appl 1997; Igumenshchev, Abramowicz,
\& Novikov 1998; Popham \& Gammie 1998). Among these papers, we feel that
the one of Narayan, Kato, \& Honma (1997, hereafter NKH97) was most clearly
written, and we would like to use it as a representative of the previous
works on global solutions of ADAF.

NKH97 concentrated itself on the global structure and dynamics of ADAF
around black holes. It did not specify any particular values for the
accretion rate and the optical depth, so the global solution obtained in it
could be applied to both the above mentioned regimes in which
advection-dominated accretion can occur. This is an advantage of that paper.
However, there are some points that we feel questionable. The first is about
the boundary conditions of the flow, which are necessary and important for
obtaining a global solution. An accretion flow into a black hole is
definitely supersonic when crossing the hole's horizon, and it is likely to
be subsonic when locating at some large distance from the hole, therefore
the flow motion must be transonic, i.e. there must exist at least one sonic
point between the horizon and the large distance. Among these three boundary
conditions what is unclear for an ADAF solution is the outer one. No one
knows yet where and how an accretion flow becomes advection-dominated, or
what kind of structure to which an ADAF should connect outwards. Perhaps
because Shakura \& Sunyaev model is the only well studied and successful one
in the accretion disk theory, NKH97, as well as almost all previous works on
global ADAF solutions, assumed that an ADAF connects outwards to a thin,
cool, Keplerian accretion disk. However, this is only an assumption, and yet
no justification has been made that it should be so. One may ask if
conditions other than the thin disk condition are also possible for ADAF's
outer boundary. What we wish to have is a complete overview of the flow
parameter space, in which all possible ADAF solutions corresponding to
different boundary conditions are presented. This is the first purpose of
our present paper.

The second question concerning NKH97, which we also try to answer in the
present paper, is about discrepancies between the results of NKH97 for
viscous ADAF and that in the literature for adiabatic, inviscid flows (e.g.
Abramowicz \& Zurek 1981; Chakrabarti 1990; Chakrabarti 1996b; Lu et al.
1997). Namely, for viscous ADAF there is always only one sonic point which
is located close to the black hole, i.e. at several gravitational radii, and
no shocks were found in the flow; while for the inviscid flow the sonic
point can locate further away from the black hole, and there may exist two
physical sonic points in the flow, consequently shocks may form. It is hard
to understand these qualitative discrepancies, because adiabatic flows are
the ideal case of ADAF, in the sense that all the energy of the accreted
matter is carried into the black hole without any loss, there should be no
such a jump between the dynamical behavior of adiabatic flows and that of
ADAF. Rather, we think it is more reasonable to expect some similarities
between the solution for adiabatic flows and that for ADAF, as well as a
continuous and smooth change in the properties of flows, i.e. from inviscid
flows via weakly viscous flows to strongly viscous flows.

\section{Dynamics}
\subsection{Equations}

We consider a steady state axisymmetric accretion flow. The governing
equations are unique, although they may be written in more or less different
forms by various authors. Here we employ exactly the same set of equations
as in NKH97, which is similar to the 'slim disk equations' of Abramowicz et
al. (1988). In this formulation the hydrostatic equilibrium in the vertical
direction is assumed, all physical variables are averaged vertically and
thus are functions only of the cylindrical radius $R$ , the following
resulting height-integrated equations, as shown by Narayan \& Yi (1995), are
a very good representation of the behavior of both thin disks and ADAF which
tend to be nearly spherical. \\ 
(i) The continuity equation:
\begin{eqnarray}
&&\frac d{dR}(2\pi R)(2H\rho \upsilon )=0       
\end{eqnarray}
or, its integration form
\begin{eqnarray}
&&-4\pi RH\rho \upsilon =\dot M=constant   
\end{eqnarray}
where $\rho \ $is the density of the accreted gas, $H$ is the vertical
half-thickness of the flow, $\upsilon $ is the radial velocity which is
defined to be negative when the flow is inward, and is the mass accretion
rate. \\ 
(ii) The equation of vertical hydrostatic equilibrium:
\begin{eqnarray}
&&H=(\frac 52)^{1/2}\frac{c_s}{\Omega _k},
\end{eqnarray}
where the coefficient (5/2)$^{1/2}$ is chosen on the basis of the
self-similar solution obtained by Narayan \& Yi (1995); $c_s$ is the
isothermal sound speed of the gas defined as
\begin{eqnarray}
&&c_s^2=p/\rho,
\end{eqnarray}
with $p$ being the pressure; and $\Omega _k$ is the Keplerian angular
velocity which takes the form
\begin{eqnarray}
&&\Omega _k^2=\frac{GM}{(R-R_g)^2R},                          
\end{eqnarray}
if the gravitational potential of the central black hole is assumed to be
described by Paczynski \& Wiita (1980) potential
\begin{eqnarray}
&&\Phi (R)=-\frac{GM}{(R-R_g)},                                 
\end{eqnarray}
with $M$ being the mass of the black hole, and $R_g$ the
gravitational radius, $R_g$ $\equiv $ $2GM/c$$^2$ . \\
(iii) The radial momentum equation:
\begin{eqnarray}
&&\upsilon \frac{d\upsilon }{dR}=\Omega ^2R-\Omega _k^2R-\frac 1\rho \frac
d{dR}(\rho c_s^2),                                             
\end{eqnarray}
where $\Omega $ is the angular velocity of the gas. \\ 
(iv) The angular momentum equation:
\begin{eqnarray}
&&\upsilon \frac d{dR}(\Omega R^2)=\frac 1{\rho RH}\frac
d{dR}(\nu \rho
R^3H\frac{d\Omega }{dR}),    
\end{eqnarray}
where $\nu $ is the kinematic coefficient of viscosity. In the spirit of
Shakura \& Sunyaev (1973), $\nu $ can be written as
\begin{eqnarray}
&&\nu =\frac{c_s^2}{\Omega _k},  
\end{eqnarray}
where $\alpha $ is assumed to be a constant. Substituting equation (9) in
equation (8) and integrating, one obtains
\begin{eqnarray}
&&\frac{d\Omega }{dR}=\frac{\upsilon \Omega _k(\Omega
R^2-j)}{\alpha R^2c_s^2},
\end{eqnarray}
where the integration constant $j$ represents the specific angular momentum
per unit mass accreted by the black hole. \\ 
(v) The energy equation for ADAF:
\begin{eqnarray}
&&\frac{\rho \upsilon }{(\gamma-1)}\frac{dc_s^2}{dR}-c_s^2\upsilon
\frac{d\rho }{dR}=\frac{\alpha \rho c_s^2R^2}{\Omega _k}(\frac{d\Omega }
{dR})^2, 
\end{eqnarray}
where $\gamma $ is the ratio of specific heats of the gas. \\
As mentioned in NKH97, Shakura \& Sunyaev (1973) originally wrote a simpler
prescription for the shear stress of the form
\begin{eqnarray}
&&shear stress = -\alpha p\frac{d\ln \Omega _k}{d\ln R}.      
\end{eqnarray}
With this prescription, equation (8) is modified to a different form which
integrates to give an algebraic relation instead of differential equation
(10), i.e. the resulting set of equations contains one less differential
equation, and the task of finding solutions becomes considerably easier. For
this reason the shear stress prescription (12) has been used by most of the
authors in the field (see Chakrabarti 1996a for references).

\subsection{Boundary conditions and numerical method}

The set of dynamical equations consist of two algebraic equations (2) and
(3), and three first-order differential equations (7), (10), and (11), for
five unknown variables $H$, $\rho$, $c_s$, $\upsilon $, and $\Omega $, all
being functions of $R$. As stated in the introduction section, a global
solution for black hole accretion requires three boundary conditions, 
namely, that at the inner boundary $R_{in}$, at the outer boundary
$R_{out}$, and at the sonic point $R_s$, respectively. We discuss them in
turn. \\
(i) The sonic point condition: \\
By combining equations (1), (3), (7), (10), and
(11) to eliminate $H$, $d\Omega/dR$, $d\ln\rho/dR$, and $dc_s/dR$, one
obtains the following differential equation for $d\upsilon /dR$:
\begin{eqnarray}
\nonumber
&&(\frac{2\gamma }{\gamma +1}-\frac{\upsilon ^2}{c_s^2})\frac{d\ln
\mid \upsilon \mid}{dR}=\frac{(\Omega _k^2-\Omega
^2)R}{c_s^2}-\frac{2\gamma }{\gamma +1}\\
\nonumber \\
&& \ \ \ \times(\frac {1}{R}-\frac{d\ln \Omega _k}{dR})+\frac{(\gamma
 -1)\Omega _k\upsilon(\Omega R^2-j)^2}{\alpha (\gamma +1)R^2c_s^4}.  
\end{eqnarray}
To make a smooth sonic transition for the flow, $d\upsilon /dR$ must be
well behaved across the sonic point, this gives two boundary conditions:
\begin{eqnarray}
&&\upsilon ^2-\frac{2\gamma }{\gamma +1}c_s^2=0, \ \ \   R=R_s,
\end{eqnarray}
\begin{eqnarray}
&&\frac{(\Omega _k^2-\Omega ^2)R}{c_s^2}-\frac{2\gamma
}{\gamma +1}(\frac 1R-
\frac{d\ln \Omega _k}{dR})+\frac{(\gamma -1)\Omega _k\upsilon (\Omega
R^2-j)^2}{\alpha (\gamma +1)R^2c_s^4}=0,  \ \ \    R=R_s.
\end{eqnarray}
Conditions (14) and (15) are exactly the same as in NKH97, and essentially
the same as in all the previous works, because a sonic point in physics is
just a critical point of ordinary differential equation (13) in mathematics,
there is no question about it. \\
(ii) The inner boundary condition: \\ 
NKH97 imposed a no-torque condition ($d\Omega /dR = 0$) at $R = R_{in}$,
which corresponds to (cf. eq. [10])
\begin{eqnarray}
&&\Omega R^2-j=0,  \ \ \ R=R_{in},
\end{eqnarray}
and argued that, although technically this condition must be applied at the
black hole horizon (i.e.$ R = R_{in} = R_g$), in practice it could be
applied equally well at any other radius between the horizon and the sonic
point, or even at the sonic point itself. The fact that such a condition
arises is because the viscous stress term in equation (8) is diffusive in
form, signals due to the shear stress can propagate backward even in the
supersonic zone of the flow, and this makes it necessary to impose a
downstream boundary condition on the angular momentum. If the prescription
for shear stress (12) is adopted as in most of the previous studies, signals
cannot propagate backward from the supersonic zone, then condition (16) is
unnecessary. \\
(iii) The outer boundary condition: \\
Unlike the above two ones, this condition is very uncertain, and it is the
treatment of this condition that makes the numerical model of global
solution. NKH97 assumed that the gas starts off from $R_{out} = 10^6R_s$
in a state that corresponds to a standard Shakura-Sunyaev thin disk, and
accordingly applied two conditions:
\begin{eqnarray}
&&\Omega =\Omega _k, \ \ \ R=R_{out},                             
\end{eqnarray}
\begin{eqnarray}
&&c_s=10^{-3}\Omega _kR,  \ \ \ R=R_{out},                       
\end{eqnarray}
of which condition (17) means that the flow is Keplerian rotating, and
condition (18) means that the disk is cool, or equivalently thin ($H\sim 
10^{-3}R$ , cf. eq. [3]). With five boundary conditions (14), (15),
(16), (17), and (18), NKH97 dealt with a very challenging numerical boundary
value problem using a relaxation method. Since there are only three
differential equations (7), (10), and (11) to be solved, the five boundary
conditions enabled them to determine the two unknowns, i.e. the sonic radius
$R_s$ and the integration constant $j$ as the eigenvalues. However, such 
an eigenvalue problem is numerically very time-consuming and difficult to
solve; besides, one may still wonder how to know the location of outer
boundary (why should it be $10^6R_s$\ ?), and whether conditions other
than (17) and (18) are also possible for ADAF solutions.

Chakrabarti and his collaborators introduced a very clever procedure (e.g.
Chakrabarti 1996a). The difficulty of finding the eigenvalues was simply
avoided: $R_s$ and $j$ were chosen to be free parameters and their values
were supplied, and then the equations were integrated from the sonic point
inwards until the black hole horizon, and outwards until the Keplerian value
of angular momentum be achieved, i.e. condition (17) be satisfied. The outer
boundary of the flow $R_{out}$ was solved out this way, rather than being
specified as in NKH97. However, it should be noted that condition (17) is
only a half of the thin disk condition, while the other half, i.e. condition
(18) which describes the shape (or temperature) of the flow at $R_{out}$, 
was not mentioned in Chakrabarti's work. Indeed, Chakrabarti's approach is
to let the outer boundary conditions automatically follow from an already
known solution, there is no guarantee that solutions constructed in such a
way will always be physically reasonable. Incorrect choices of eigenvalues
may cause troubled solutions with which no astrophysically acceptable
outer boundary conditions are consistent. Very recently, adopting a
procedure similar to Chakrabarti's, Igumenshchev, Abramowicz, \& Novikov
(1998) obtained global ADAF solutions for which the flow's shape is slim
everywhere ($H/R \leq 1$) and the distribution of angular momentum has a
super-Keplerian part at the outer boundary. In particular, they found that
ADAF solution could match outwards to the solution of Shapiro, Lightman, \&
Eardley (1976) in the case of bremsstrahlung cooling. The importance of
these results is that outer boundary conditions other than the thin disk
conditions (17) and (18) are shown to be possible for ADAF solutions.

In the present paper we solve the dynamical equations described in 2.1 in a
way also similar to Chakrabarti's. There are three differential equations
(7), (10), and (11) for three unknown variables $c_s$, $\upsilon $, and 
$\Omega $ (the term $d\ln\rho /dR$ in eqs. [7] and [11] can be substituted
for by using eqs. [1] and [3]). We integrate the three equations from the
sonic point both inwards and outwards, and do not specify any ad hoc outer
boundary conditions. To do so, we first take the sonic radius $R_s$ and
the integration constant $j$ as two free parameters. Three more boundary
values, namely the values of $c_s$, $\upsilon $, and $\Omega $ at $R =
R_s$ are needed to start the integration, but there exist two constraint
conditions (14) and (15), thus only one more boundary value has to be
supplied. Our choice is to apply no-torque condition (16) at the sonic
point $R_s$, this means that the value of $\Omega $ at $R_s$ is supplied,
then the values of $c_s$ and $\upsilon $ at $R_s$ can be obtained from
conditions (14) and (15). We use the standard fourth-order Runge-Kutta
method to perform the integration from $R_s$. The derivative $d\upsilon /
dR$ at $R_s$ is evaluated by applying l'H\^opital's rule and solving a
quadratic equation. The inward part of the solution extends until the
black hole horizon, showing the flow's supersonic motion; and the outward,
subsonic part extends until a large radial distance which can be
considered to be the flow's outer boundary. After having $c_s$,$\upsilon $
, and $\Omega $ as functions of $R$, other variables $H$, $\rho $, and $p$
can be calculated from equations (2), (3), and (4), and a global solution
is obtained. By varying the values of two free parameters $R_s$ and $j$,
we can find out all the possibilities of solutions: although some of the
solutions constructed this way may not be physically acceptable, no
physical solutions will be missed under such a 'carpet bombing'.

\section{Results and discussion}

We now present our numerical results. In our calculations the ratio of
specific heats of the accreted gas $\gamma $ is fixed to be equal to 1.5.
Figure 1 is an overview of the results which shows all the possibilities of
solutions in the flow parameter space spanned by the sonic radius $R_s$
and the specific angular momentum accreted by the black hole $j$. The
figure is for the viscosity parameter $\alpha $ = 0.01. When $\alpha $ is
taken to be other non-zero values the results keep qualitatively similar
and change only quantitatively. It is seen from the figure that the whole
parameter space is divided into six regions. The dashed line is for
Keplerian angular momentum ($l_k = \Omega _kR^2$) corresponding to given
$R_s$, and region I is that above the dashed line. This region is
unphysical, because it would require the square of radial velocity
$\upsilon ^2$ at the sonic point to have negative values; or in other
words, the value of $j$ is chosen to form too high a centrifugal barrier,
so that it is impossible for the gas to be accreted. Region II is that
above the dotted line which is also unphysical, again because of wrong
choices of the parameters. In this region the derivative $d\upsilon /dR$
at $R_s$, calculated using l'H\^{o}pital's rule, has no real solution,
such a sonic point is classified as of spiral type and is not physical.
Apart from these two no-solution regions, the other regions in the
parameter space all give solutions. We describe them in detail in the
following subsections.

\subsection{ADAF-thin disk solutions}

We obtain ADAF solutions that connect outwards to thin disk solutions, i.e.
conditions (17) and (18) at the outer boundary $R_{out}$ are both
fulfilled. These solutions are just what constructed in NKH97, and all sit
on the solid line AB in figure 1, which forms a part of the boundary
between regions III and V. Figures 2a-c give a typical example of this
class of solution, for which $R_s = 2.29R_g$, and $j = 1.749 (cR_g)$.
Figure 2a shows the radial velocity $\upsilon $ and the sound speed 
$c_s$ of the flow as functions of the radius $R$, drawn as a solid line
and a dashed line, respectively. The crossing point of the two lines
(marked by a filled circle) is the sonic point $R_s$. In figure 2b the
radial distribution of the angular momentum of the flow ($l = \Omega R^2$)
is shown by a solid line, and the dashed line gives the Keplerian angular
momentum distribution ($l_k = \Omega _kR^2$). It is seen that beyond the
outer boundary $R_{out} = 10^{3.99}R_g$ (marked by a filled square) where
the flow's angular momentum equals the Keplerian value (condition [17]),
there is a narrow region for which the rotating motion of the flow is
super-Keplerian. This feature was not found in NKH97, but has been noticed
recently in Igumenshchev, Abramowicz, \& Novikov (1998). We are agreed
with the latter authors about the physical reason of this feature: in
this narrow region the pressure increases with radius, which is due to the
rapid decrease of the disk thickness (cf. Fig. 2c) and the corresponding
increase in the matter density, the fact that both the gravitational force
and the pressure-gradient force are directed inwards results in
super-Keplerian rotation. Indeed, super-Keplerian rotation often occurs in a
region where a sharp transition in qualitative properties of the accretion
flow takes place, e.g. at the cusp-like inner edge of thick disks
(Abramowicz, Calvani, \& Nobili 1980). NKH97 also noticed the 
super-Keplerian rotation, but on the inside, i.e. in a narrow region just
outside the sonic point for flows with low values of $\alpha $. However, the
way in which we find the super-Keplerian rotation in the outermost region of
ADAF is different from that of Igumenshchev, Abramowicz, \& Novikov (1998).
Those authors still assumed some outer boundary condition, i.e. at 
$R_{out}$ the angular velocity of the flow was fixed as a fraction of
the Keplerian angular velocity, although they showed that the presence of
super-Keplerian rotation did not depend on the particular value taken for
$\Omega (R_{out})$; we do not assume any outer boundary conditions, the
super-Keplerian rotation is found naturally by the outward integration of
the equations. Figure 2c shows the relative thickness of the disk $H/R$ 
for all the range of radii. It is seen that the disk is thin in its inner
region, is slim in the middle region ($H/R$ is close to 1), and that the
thickness of the disk decreases rapidly with radius in the outer region.
Beyond the outer boundary $R_{out}$ (the filled square) where the thin
disk condition (18) is fulfilled, $H$ tends to be zero, and the outward
integration has to be stopped.

It should be addressed that the way of finding ADAF-thin disk solutions in
NKH97 is very restrictive. Adopting five boundary conditions (14)-(18) for
three differential equations (7), (10), and (11) enabled those authors to
determine two unknowns $R_s$ and $j$ as the eigenvalues. However, the range of
variation for the value of $R_{out}$ is very large (cf. Fig. 3), perhaps
because of this uncertainty NKH97 constructed only a few global solutions
with $R_{out} = 10^6R_s$. We take $R_s$ and $j$ as free parameters instead,
then $R_{out}$ is found out by the meeting of conditions (17) and (18) in
the outward integration of the equations, rather than being assumed.
Technically, our approach is equivalent to that of NKH97 because the three
quantities $R_s$, $j$, and $R_{out}$ in ADAF-thin disk solutions are the 
eigenvalue of each other: if one of them is correctly given, then the
other two are fixed accordingly and have to be found out. Figure 1 has
shown such an unique correspondence between $R_s$ and $j$ (the solid line
AB). In figure 3 we present the unique correspondence between $R_s$ and
$R_{out}$ for three different values of $\alpha $, the middle line AB is
for $\alpha  = 0.01$ which corresponds to the solid line AB in figure 1.
It is obvious from figures 1 and 3 that the acceptable range of $R_{out}$
is very large, while those of $R_s$ and $j$ are very small. Thus our
approach has an advantage that, by varying the values of $R_s$ and $j$, we
are able to find all the ADAF-thin disk solutions, not only some examples
as in NKH97.

\subsection{ADAF-thick disk solutions}

For parameters $R_s$ and $j$ located in region V of figure 1, i.e. the left and
the lower part of the parameter space, global solutions are of another
class: an ADAF solution connects outwards to a thick disk solution. Figures
4a-c provide an example of this class of solution, for which $R_s = 2.29 R
_g$, $j = 1.74 (cR_g)$. Figure 4a shows the radial variation of the radial
velocity $\upsilon $ (solid line) and of the sound speed $c_s$ (dashed
line), the crossing point (the filled circle) of the two lines is the sonic
point $R_s$. Figure 4b draws the radial distribution of the flow's angular
momentum $l$ (solid line) and that of Keplerian angular momentum $l_k$
(dashed line). The crossing point (the filled square) of the two lines
defines the outer boundary of ADAF solution $R_{out} = 10^{1.837}R_g$ at
which condition (17) is fulfilled. Super-Keplerian rotation is also
presented outside $R_{out}$, also because the pressure-gradient is
directed inwards. However, the increase of pressure with radius is not due
to the sharp decrease of the disk's thickness and the corresponding
increase in the matter density as for ADAF-thin disk solutions, rather, it
is due to the increase of sound speed (i.e. temperature) as reflected by
the increase of the disk's thickness (cf. Fig. 4c and eq. [3]).

The most noticeable feature of this class of solution is that, as seen from
figure 4c, condition (18) is not fulfilled, i.e. the disk is not thin at the
outer boundary of ADAF solution $R_{out}$ (the filled square), and outside
$R_{out}$ the relative thickness $H/R$ increases with radius and gets
larger than 1. We note that, unlike NKH97, in many previous works which
assumed hydrostatic equilibrium in the vertical direction of the flow,
only condition (17), i.e. Keplerian rotating was adopted to determine the
outer boundary $R_{out}$, while the shape (or the temperature) of the flow
at $R_{out}$ was not reported (e.g. Chakrabarti 1996a and references
therein). Igumenshchev, Abramowicz, \& Novikov (1998) ruled out solutions
with $H/R >> 1$ for the whole range of radii, $R_{in} < R < R_{out}$,
because those solutions did not satisfy the slim disk condition $H/R < 1$
(Note however that those solutions are different from ours, in our
solutions the disk is thin or slim for $R < R_{out}$, and $H/R > 1$ for
$R \geq  R_{out}$ only). To our knowledge, there have been only two
papers, namely Peitz \& Appl (1997, see its Figs. 5-10) and Popham \&
Gammie (1998, see its Figs. 1-4) that assumed the flow to be in vertical
hydrostatic equilibrium and obtained results similar to ours: $H/R$
increases with $R$ and exceeds unity for large $R$. We think that from
mathematical point of view the problem of the flow in vertical hydrostatic
equilibrium with $H/R \geq 1$ needs a more sophisticated study; from
physical point of view, however, the possibility of connection of an ADAF
to a geometrically thick flow is worth considering. For example, Narayan
et al. (1998) applied an ADAF model to the Galactic center. The model
naturally accounted for the apparently contradictory observations of Sgr
A*, i.e. a moderate mass accretion rate and an extremely low bolometric
luminosity. On the other hand, Melia (1992) showed that the accretion is
likely initiated by a bow shock that dissipates most of the directed
kinetic energy and heats the gas to a temperature as high as about $7
\times 10^6$ K. Such an initial state of the accretion flow corresponds
hardly to a cool, thin disk, while a hot, thick disk is probably a better
alternative. It is also known that the model for quiescent soft X-ray
transients by Narayan, McClintock, \& Yi (1996), in which an ADAF is
surrounded by an outer geometrically thin and optically thick disk, although
being very promising, encountered some difficulties. In particular, Lasota,
Narayan, \& Yi (1996) pointed out that the outer disk is too cold to account
for the observed UV flux, and suggested that an optically thin disk could
have higher temperature. However, an optically thin and geometrically thin
disk, known as the solution of Shapiro, Lightman, \& Eardley (1976), is
thermally unstable. Therefore, an optically thin, geometrically slim or
thick disk could again be an alternative for the outer disk of Narayan,
McClintock, \& Yi model.

\subsection{$\alpha $-type solutions}

Following Abramowicz \& Chakrabarti (1990), a solution is called X type if
it is complete, i.e. if it is able to join the black hole horizon to a large
radius which can be considered as the outer boundary; while a solution is
called $\alpha $ type if it is incomplete, i.e. if it can extend to the
black hole horizon only, or to the outer boundary only. The ADAF-thin disk
solution and the ADAF- thick disk solution described in the above two
subsections are both of X type. In the $R_s-j$ parameter space (Fig. 1)
there are two regions in which solutions are of $\alpha $ type, namely
region III (bounded by the solid line AB, the dashed line, the dotted line,
and the dot-dot-dot-dashed line) and region IV (between the dotted line and
the dot-dashed line). An example of the solution in region III, for which R$
_s$ = 2.29 R$_g$, j = 1,76 (cR$_g$), is given by figures 5a-c. Figure 5a is
again for $\upsilon $ vs. $R$ (solid line) and $c_s$ vs. $R$ (dashed
line). The crossing point (the filled circle) of the two lines is the
sonic point $R_s$. The inward part of the solution extends until the black
hole horizon. In the outward integration the two lines meet again at $R =
10^{2.02}R_g$ (marked by a filled triangle). However it is not another
physical sonic point, it is a singular point of differential equation
(13), i.e. only condition (14) is satisfied there while condition (15) is
not, the derivative $d\upsilon /dR$ tends to be infinite, and the outward
integration cannot be performed any further. Such a solution is obviously
of $\alpha $ type. Figure 5b shows that the angular momentum $l$ (solid
line) is sub-Keplerian for the entire range of solution (the dashed line
is for Keplerian angular momentum $l_k$); and figure 5c tells that the
relative thickness $H/R$ keeps being smaller than 1.

The $\alpha $-type solution in region IV is somewhat different. Figures 6a-c
provide an example with $R_s = 30 R_g$, $j = 1.4 (cR_g)$. The solution
extends to $R_{out} = 10^{3.736}R_g$ (the crossing point of the solid
line for $l$ and the dashed line for $l_k$ in Fig. 6b, marked by a filled
square) defined by condition (17), but not to the black hole horizon, just
opposite to the solution in region III. After passing through the sonic
point $R_s$ (the crossing point of the solid line for $\upsilon $ and the
dashed line for $c_s$ in Fig. 6a, marked by a filled circle), the inward
integration has to be stopped at $R = 4.49 R_g$ (the other meeting point
of the two lines in Fig. 6a, marked by a filled triangle), also because of
the singularity, i.e. an infinite derivative $d\upsilon /dR$ resulting
from the fact that only condition (14) is satisfied and condition (15) is
not. The distribution of angular momentum has a super-Keplerian part
outside $R_{out}$ (Fig. 6b). The relative thickness $H/R$ increases
rapidly with radius and gets much larger than 1 for large $R$ (Fig. 6c).

A single $\alpha $-type solution is unacceptable, simply because it is not a
global solution. We present this class of solution here for the following
reasons. First, it does occupy certain regions in the $R_s-j$ parameter
space, if it were ignored then the picture of solution distribution in
parameter space would be incomplete. Second, perhaps more importantly, $
\alpha $-type solutions are not totally useless. An $\alpha $-type solution
could join another $\alpha $-type or an X-type solution via a standing
shock, forming a global solution. Lu \& Yuan (1998) obtained three types of
shock- included global solutions for adiabatic black hole accretion flows,
namely $\alpha $-X, X-$\alpha $, and $\alpha $-$\alpha $ type. We are
currently working on the problem of shock formation in ADAF around black
holes. Our initial results show that an ADAF-thick disk solution (X type) in
region V of figure 1 may join an $\alpha $-type solution in region III via a
standing shock, forming an $\alpha $-X type shock-included global solution.
However, the parameter range of $\alpha $-type solutions which imply the
existence of shocks is very small, much smaller than region III itself. For
ADAF-thin disk solutions (solid line AB in figure 1) there are indeed no
standing shocks at all, i.e. the conclusion made in NKH97 on this point is
correct. We shall discuss the problem of shocks in ADAF in detail in a
subsequent paper.

\subsection{Solutions for inviscid flows}

For comparison we show in figure 7 the distribution of solutions for
inviscid flows (the viscosity parameter $\alpha = 0$) in the $R_s-j$
parameter space. The parameter $j$ is now understood as the constant
specific angular momentum of the flow. The other parameter $R_s$ could be
equivalently replaced by the constant specific total energy if the flow is
further assumed to be adiabatic. It is seen that figure 7 is qualitatively
similar to figure 1. The whole parameter space is divided into five
regions. No solution exists in region I because a too high centrifugal
barrier prevents the flow from accreting; nor in region II because the
sonic point is of vortex type (not spiral type) which is also unphysical.
Region V is for thick disk solutions, and regions III and IV are for
$\alpha $-type solutions. However in these solutions no outer boundary
defined by condition (17) can be found: for large radii the flow's angular
momentum is always sub-Keplerian. A remarkable feature of inviscid flows
is that for these flows ADAF-thin disk solutions of NKH97 do not exist at
all. In fact, as the value of $\alpha $ decreases, the range of parameters
for ADAF-thin disk solutions shrinks, i.e. the solid line AB in figure 1
shortens. When $\alpha = 0$ (figure 7), the line AB becomes a dot of
$R_s = 2R_g$ and $j = 2(cR_g)$, the corresponding $R_{out}$ does not
exist, in other words, it becomes infinity. All these results agree with
our expectation that there should be similarities between the dynamical
behavior of ADAF and that of adiabatic inviscid flows, and that there
should be a continuous change from the behavior of viscous flows to that
of inviscid flows, i.e. when the value of $\alpha $ decreases 
continuously.

\section{Summary}

In this paper we construct global solutions describing ADAF around black
holes. The major points we wish to stress are as follows. (i) We recover the
ADAF-thin disk solution as a whole class, not only some examples as in
NKH97. (ii) We obtain a new class of global solution, namely the ADAF-thick
disk solution, and suggest possible applications of this class of solution
to some astrophysical systems. (iii) We present the $\alpha $-type solution
as still another class of solution. Most of $\alpha $-type solutions are
unacceptable because they are not global solutions, but some of them could
be involved as part of global solutions with shocks. (iv) We support the
conclusion of NKH97 that no standing shocks are relevant to ADAF-thin disk
solutions. Global solutions with shocks, if they ever exist, could only
result from combination of $\alpha $-type solutions and ADAF-thick disk
solutions. (v) We show the solution distribution for inviscid flows in the
parameter space, which agrees with the view that there should be some
similarities between ADAF and adiabatic flows, as well as some continuity
from the properties of viscous flows to those of inviscid ones.

\acknowledgments
We thank S. Kato and S. Mineshige for discussions, and R. Narayan and E.
Quataert for showing us their numerical code which enabled us to understand
the method and results of NKH97. We also thank M.A. Abramowicz for helpful
suggestions to improve our manuscript.

\newpage

\figcaption[f1.ps]{Parameter space spanned by the sonic radius $R_s$ and
the integration constant in angular momentum equation $j$ is divided into
six regions. For regions I and II there exist no solutions for ADAF. The
solid line AB is for ADAF-thin disk solutions. Regions III and IV are for
$\alpha $-type solutions. Region V is for ADAF-thick disk solutions. The
figure is for the viscosity parameter $\alpha = 0.01$, and the ratio of
specific heats $\gamma = 1.5$. See text for details. \label{fig-1}}

\figcaption[f2a.ps,f2b.ps,f2c.ps]{An example of ADAF-thin disk solutions.
In (a) the solid line shows the variation of the radial velocity $\upsilon
$ as a function of radius $R$, and the dashed line shows the variation of
the sound speed $c_s$. In (b) the solid line shows the radial variation of
the specific angular momentum $l$, and the dashed line shows the Keplerian
specific angular momentum $l_k$. (c) shows the radial variation of the
relative thickness of the flow $H/R$. The filled circle denotes $R_s$, and
the filled square denotes the outer boundary of ADAF $R_{out}$.
\label{fig-2}}

\figcaption[f3.ps]{Unique correspondence between $R_s$ and $R_{out}$ of
ADAF-thin disk solutions for three values of $\alpha $. The line AB
corresponds to the solid line AB in Fig. 1. \label{fig-3}}

\figcaption[f4a.ps,f4b.ps,f4c.ps]{An example of ADAF-thick disk
solutions. The meaning of lines and symbols is the same as in Fig. 2.
\label{fig-4}}

\figcaption[f5a.ps,f5b.ps,f5c.ps]{An example of $\alpha $-type solutions
in region III of Fig. 1. The meaning of lines and filled circle is the
same as in Fig. 2, the filled triangle denotes the singular point of
differential equation (13). \label{fig-5}}

\figcaption[f6a.ps,f6b.ps,f6c.ps]{An example of $\alpha $-type solutions
in region IV of Fig. 1. The meaning of lines and symbols is the same as in
Figs. 2 and 5. \label{fig-6}}

\figcaption[f7.ps]{Distribution of solutions for inviscid flows in $R_s$-j
parameter space. Regions I - V are similar to those of Fig. 1,
respectively, but there are no thin disk solutions here. \label{fig-7}}

\newpage

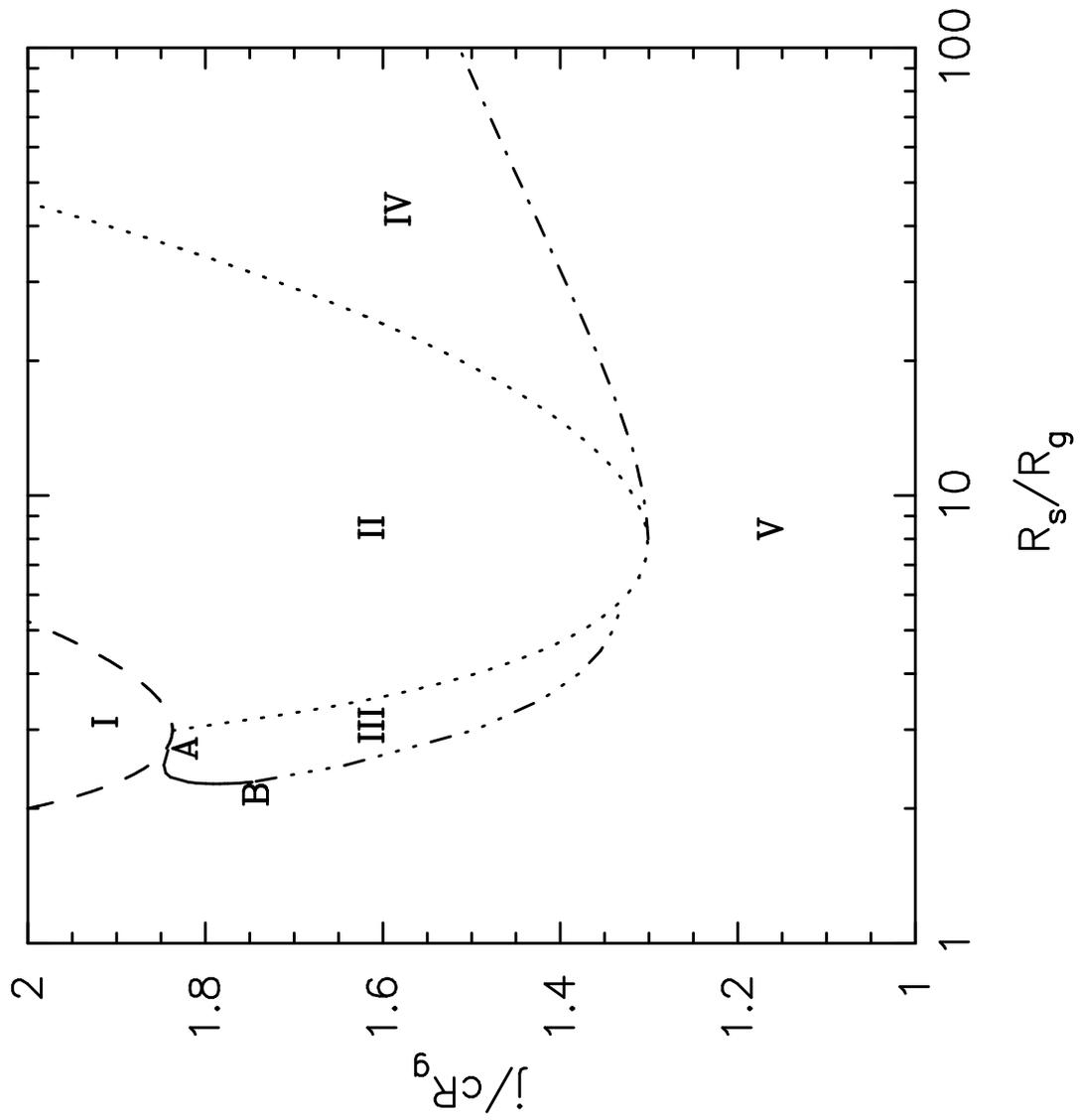
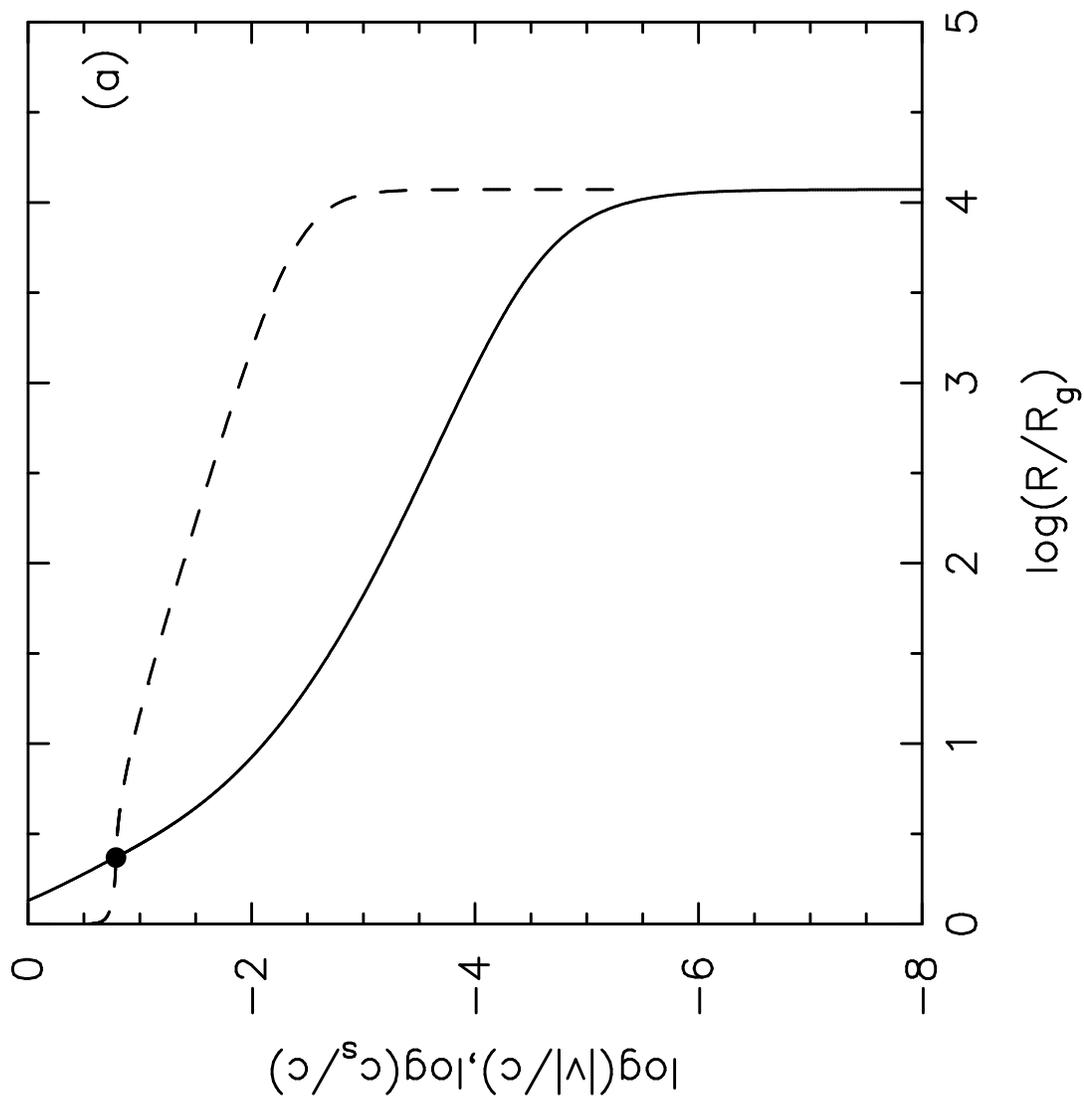
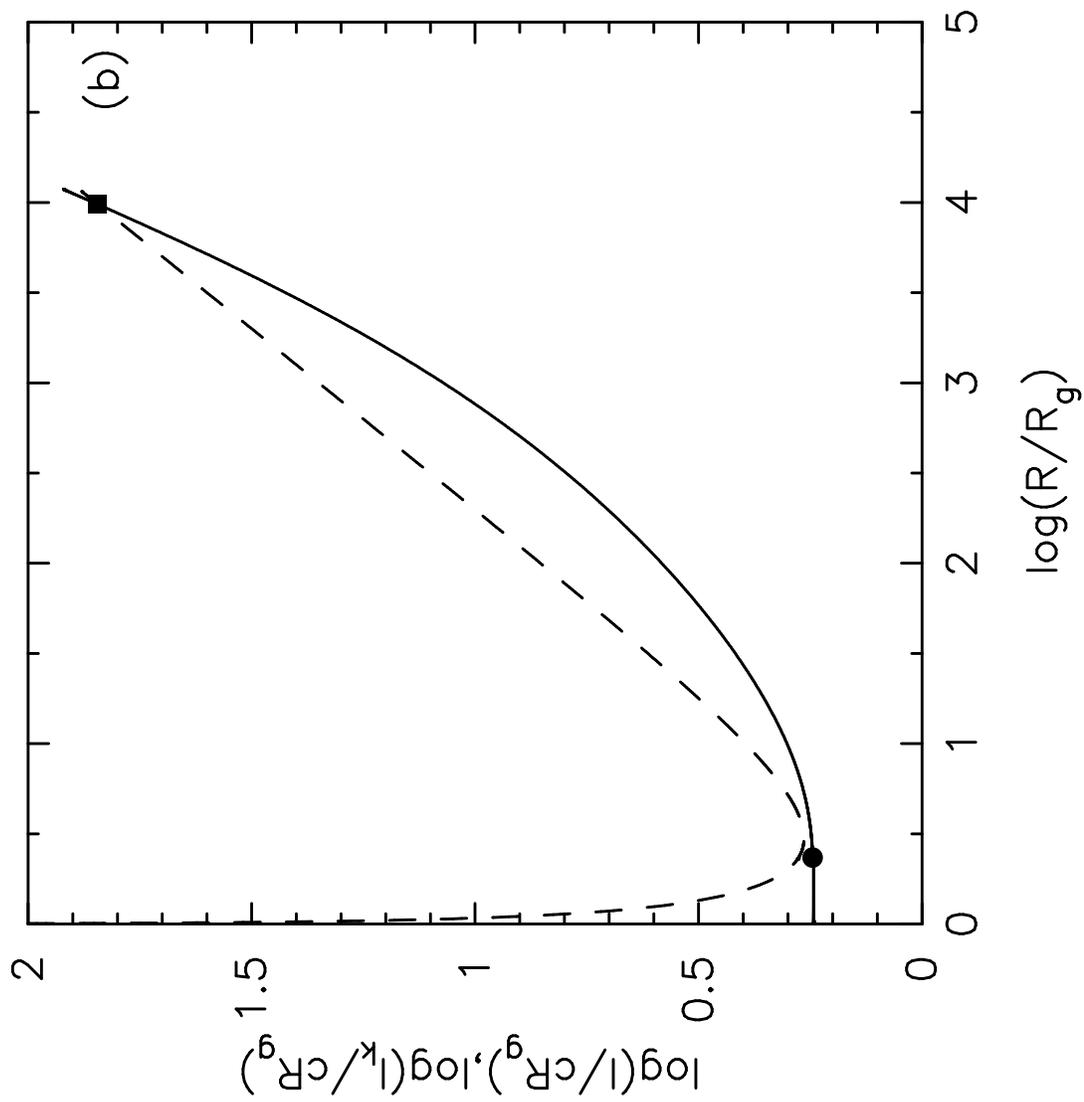
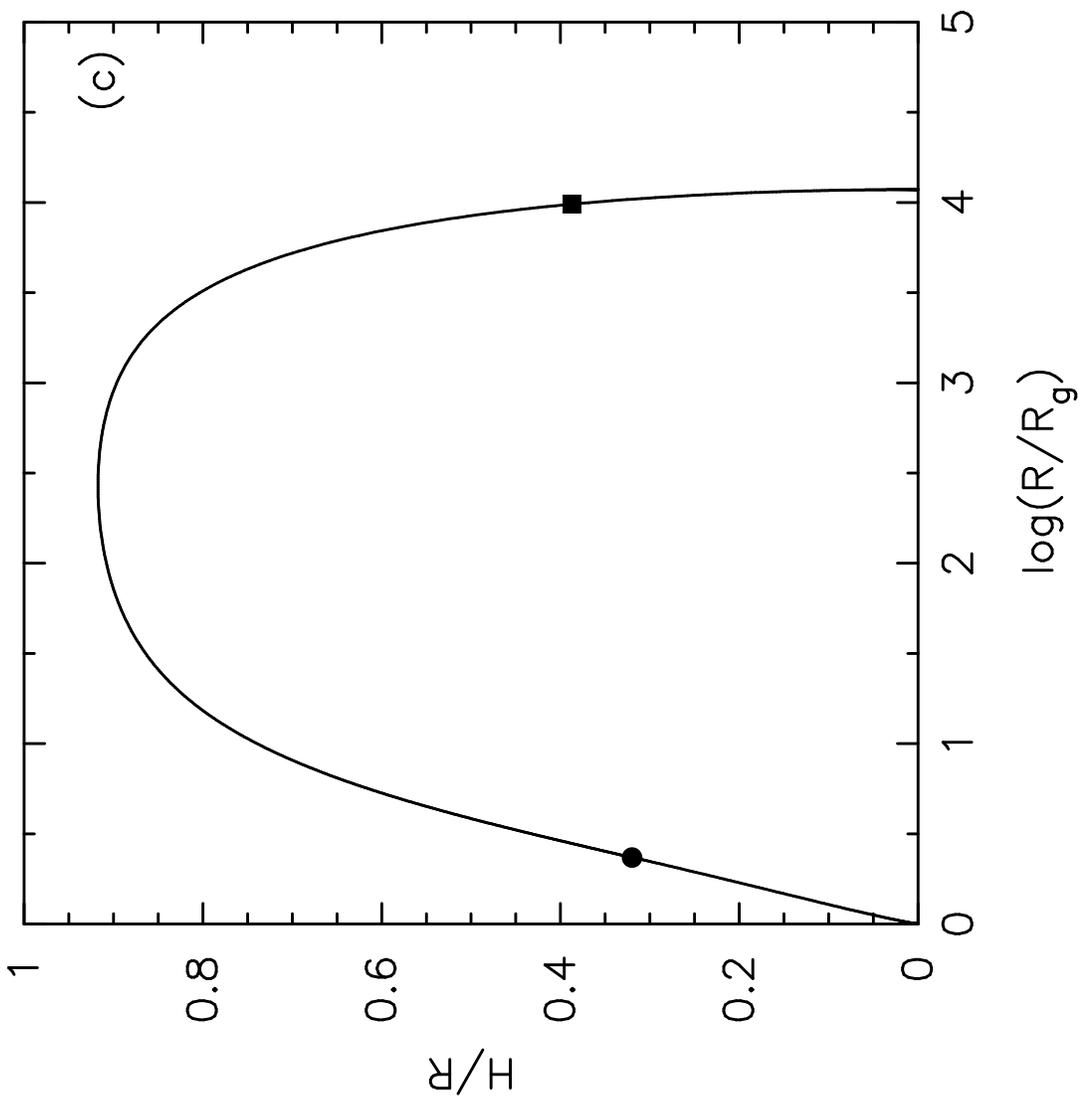
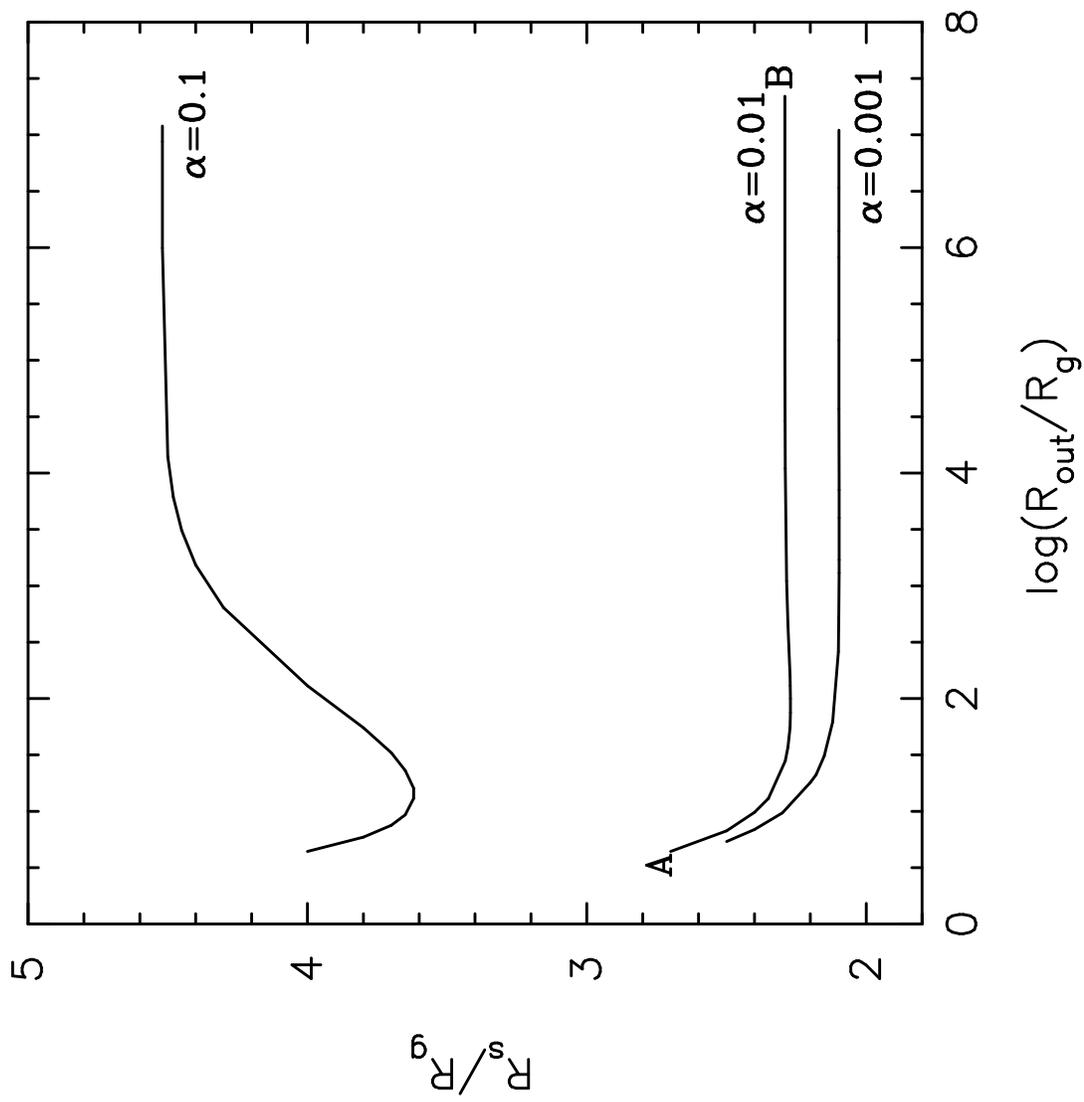
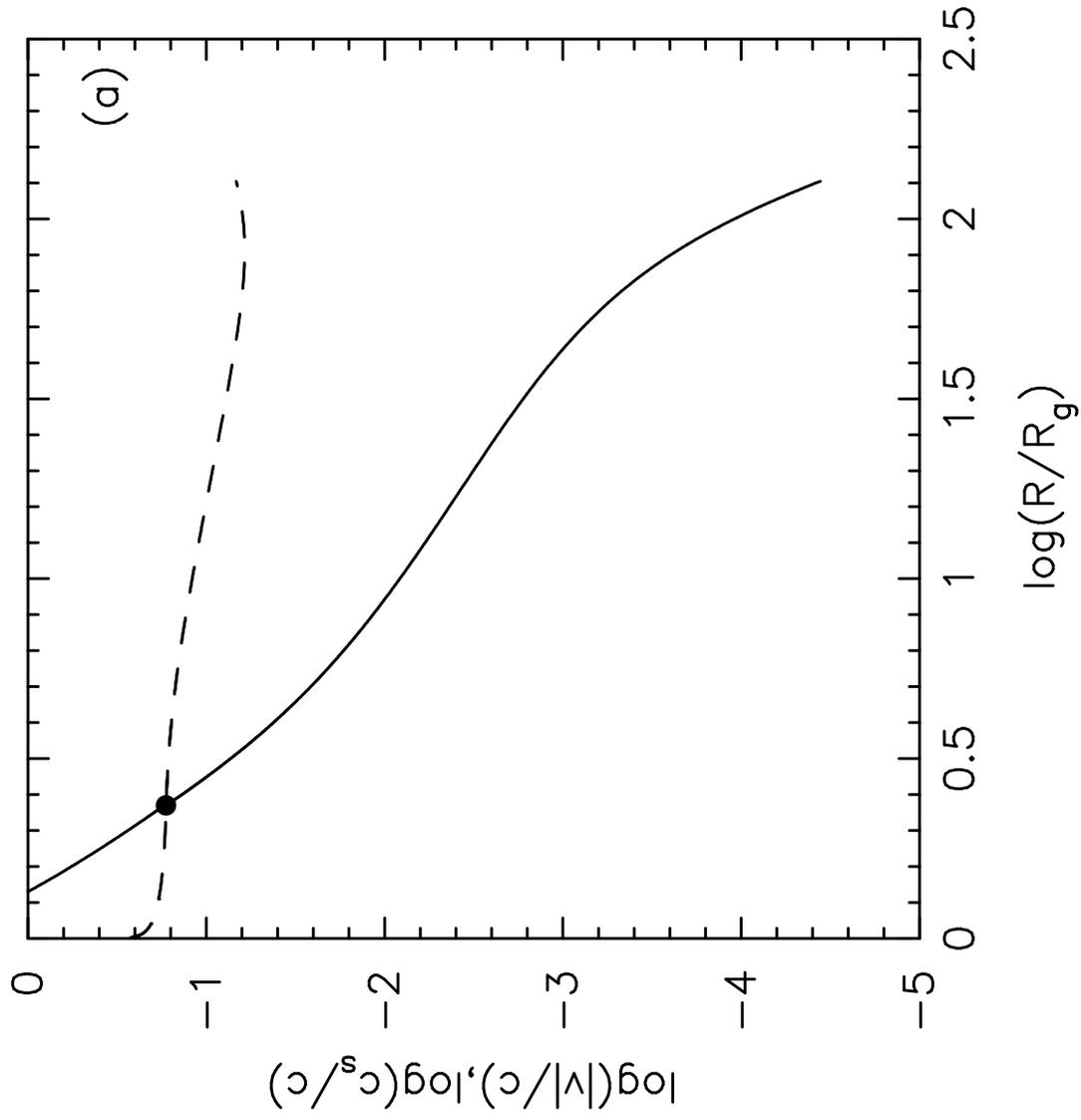
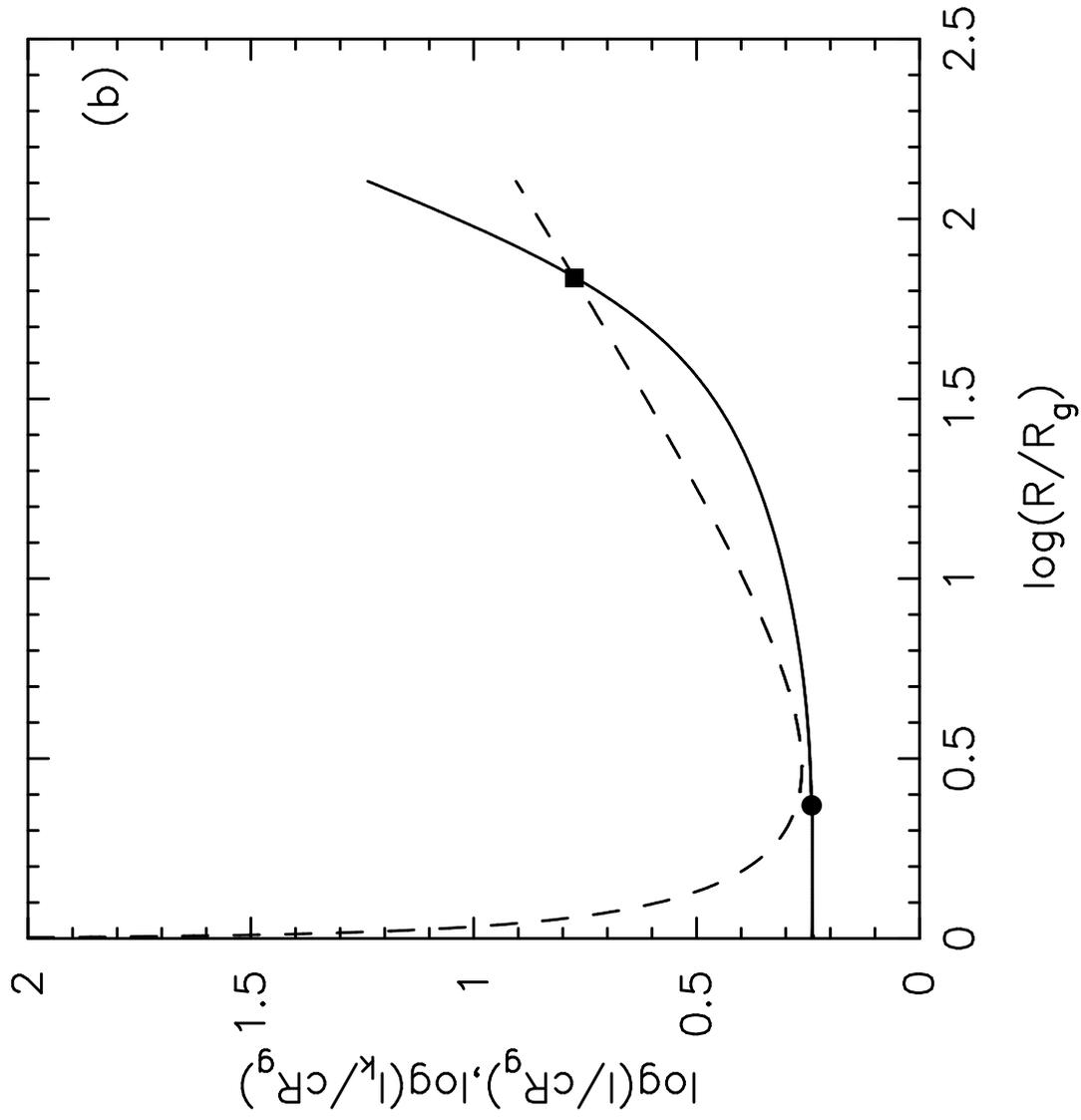
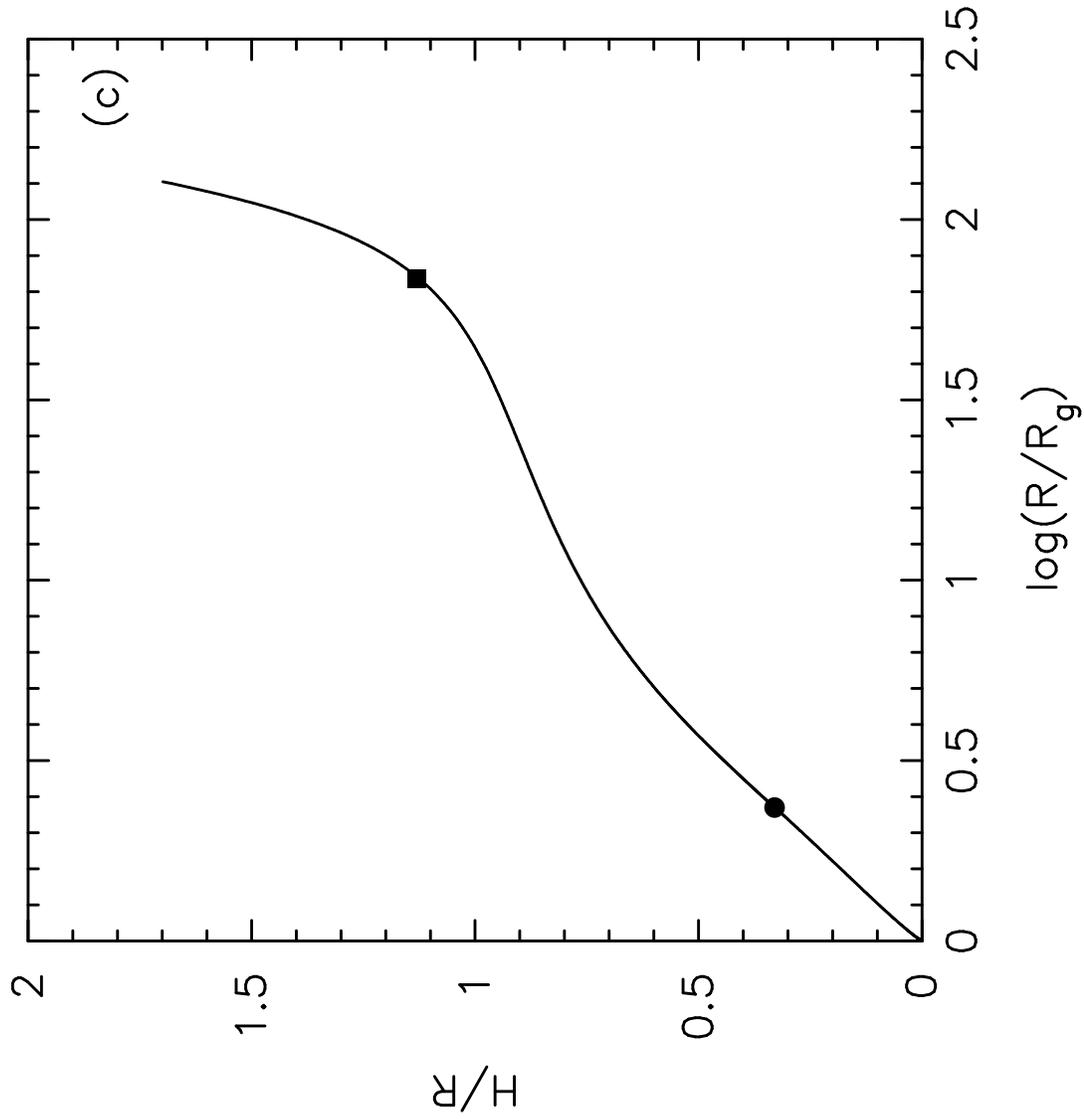
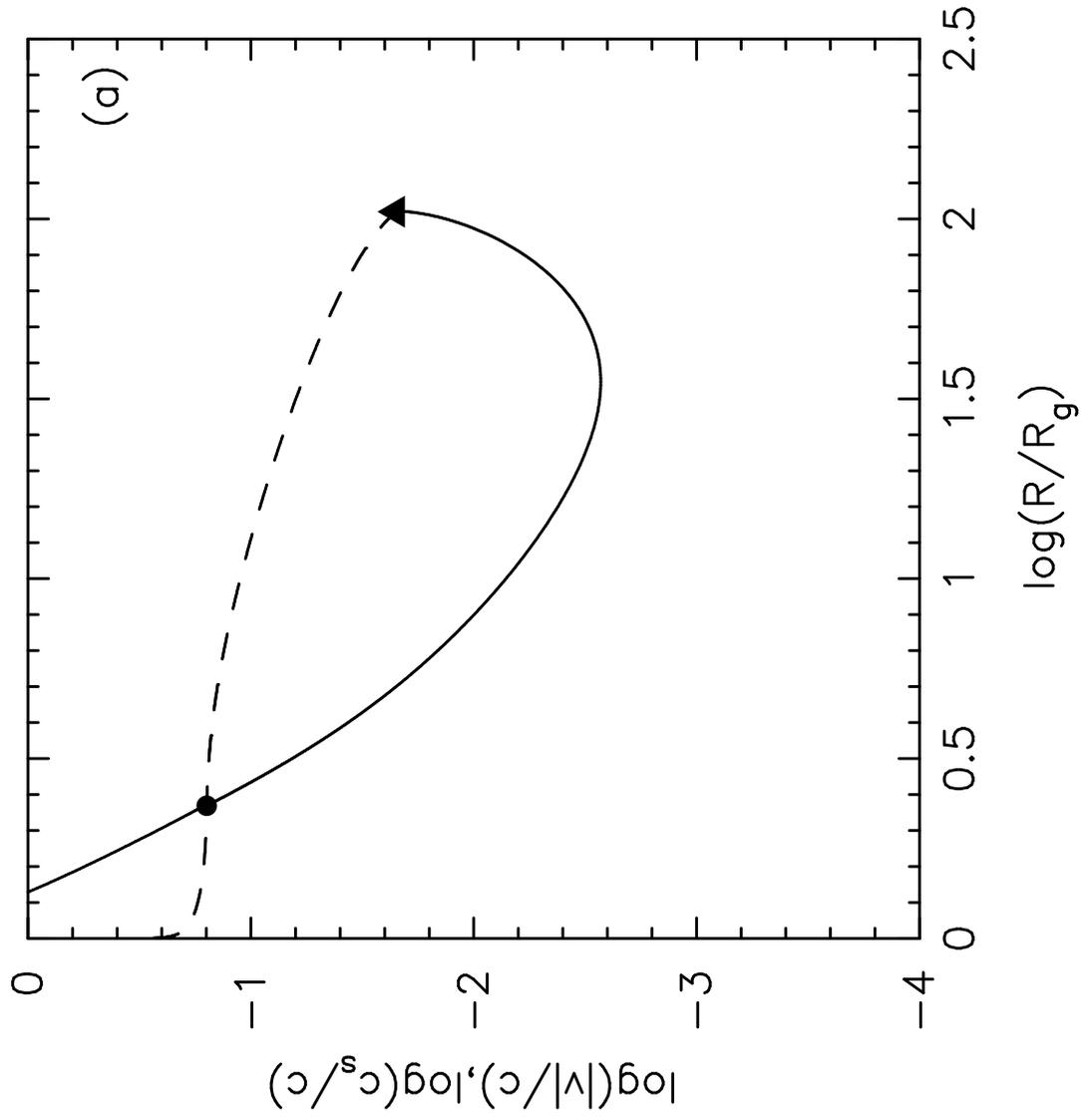
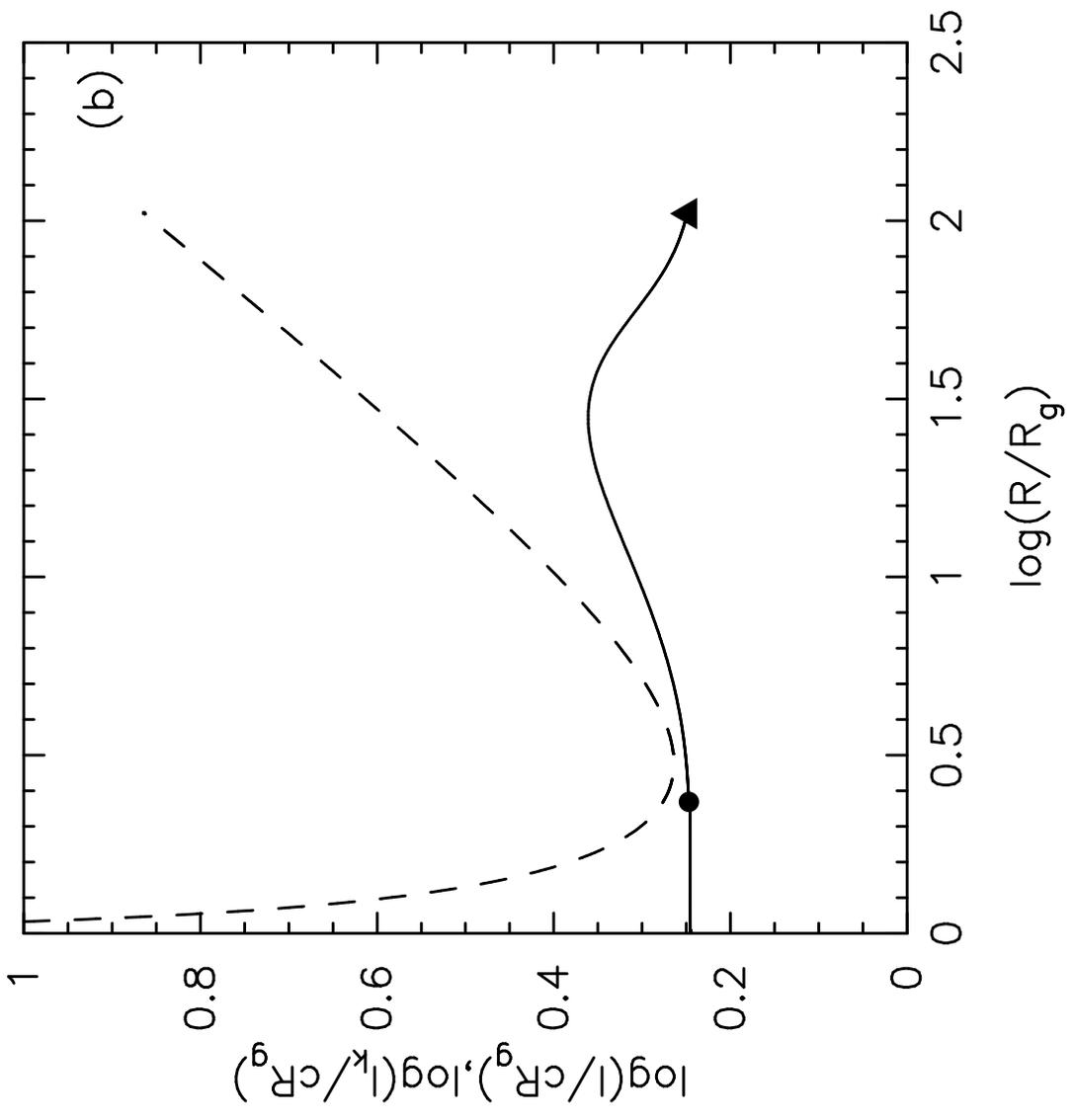
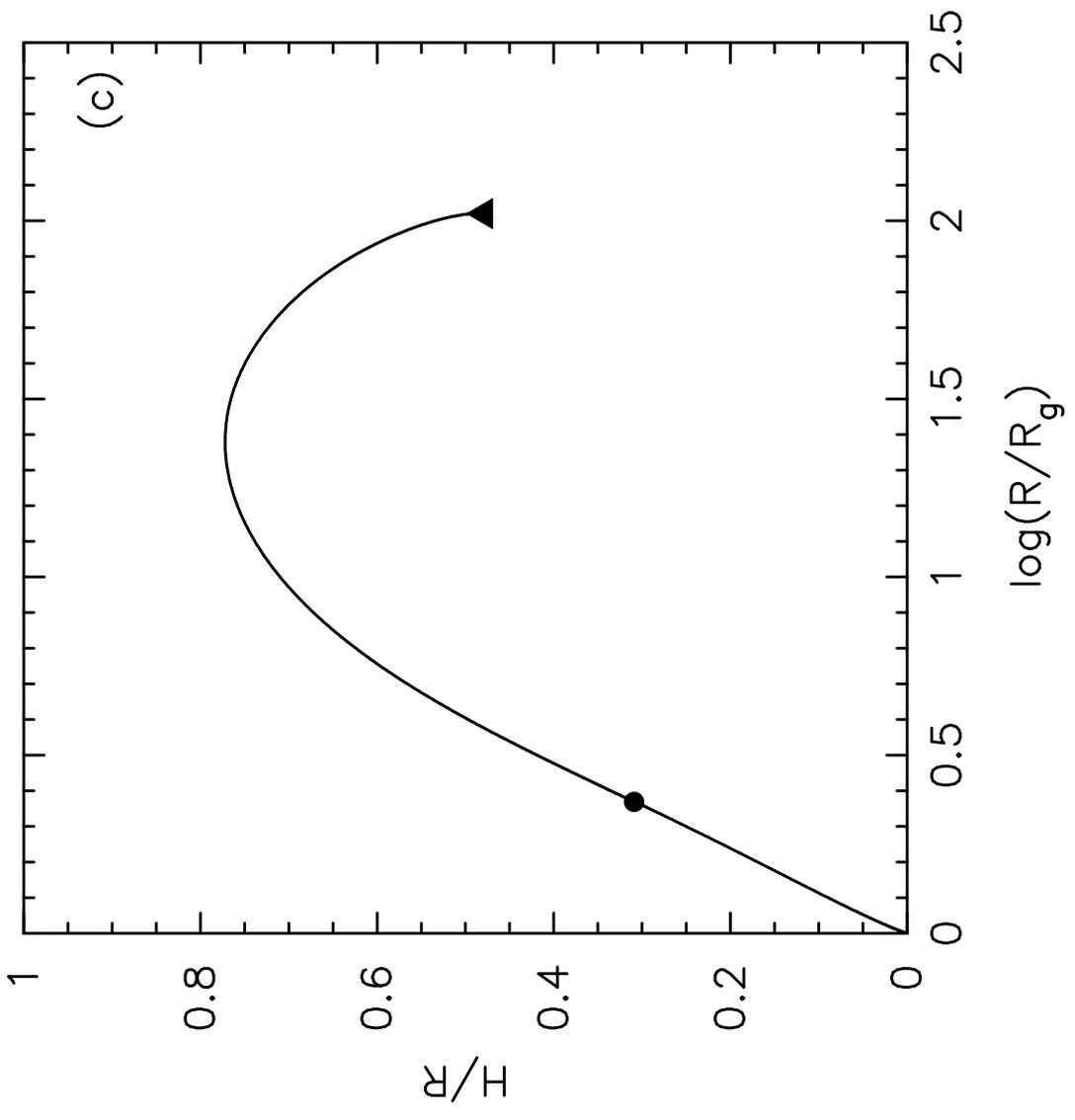
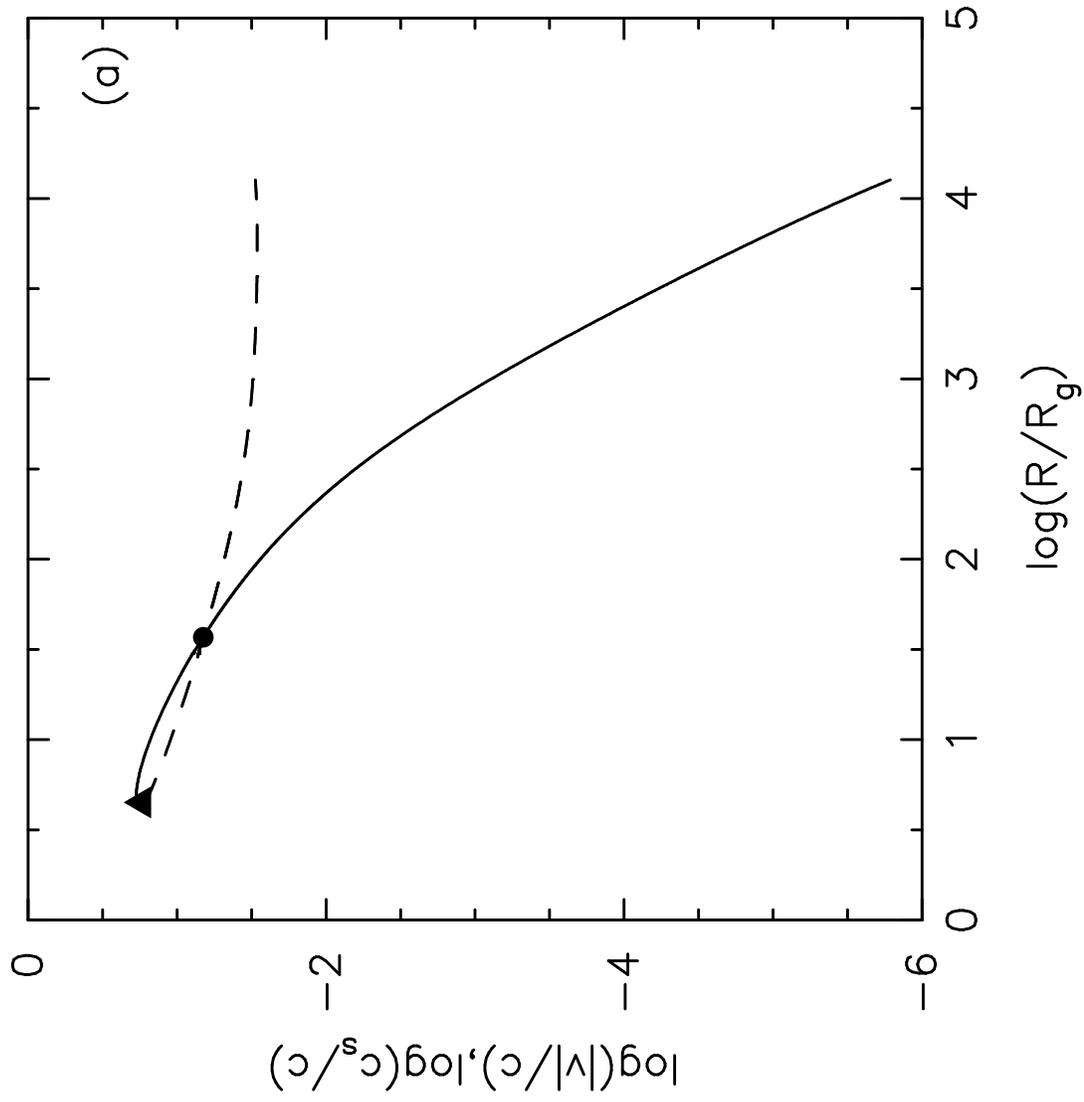
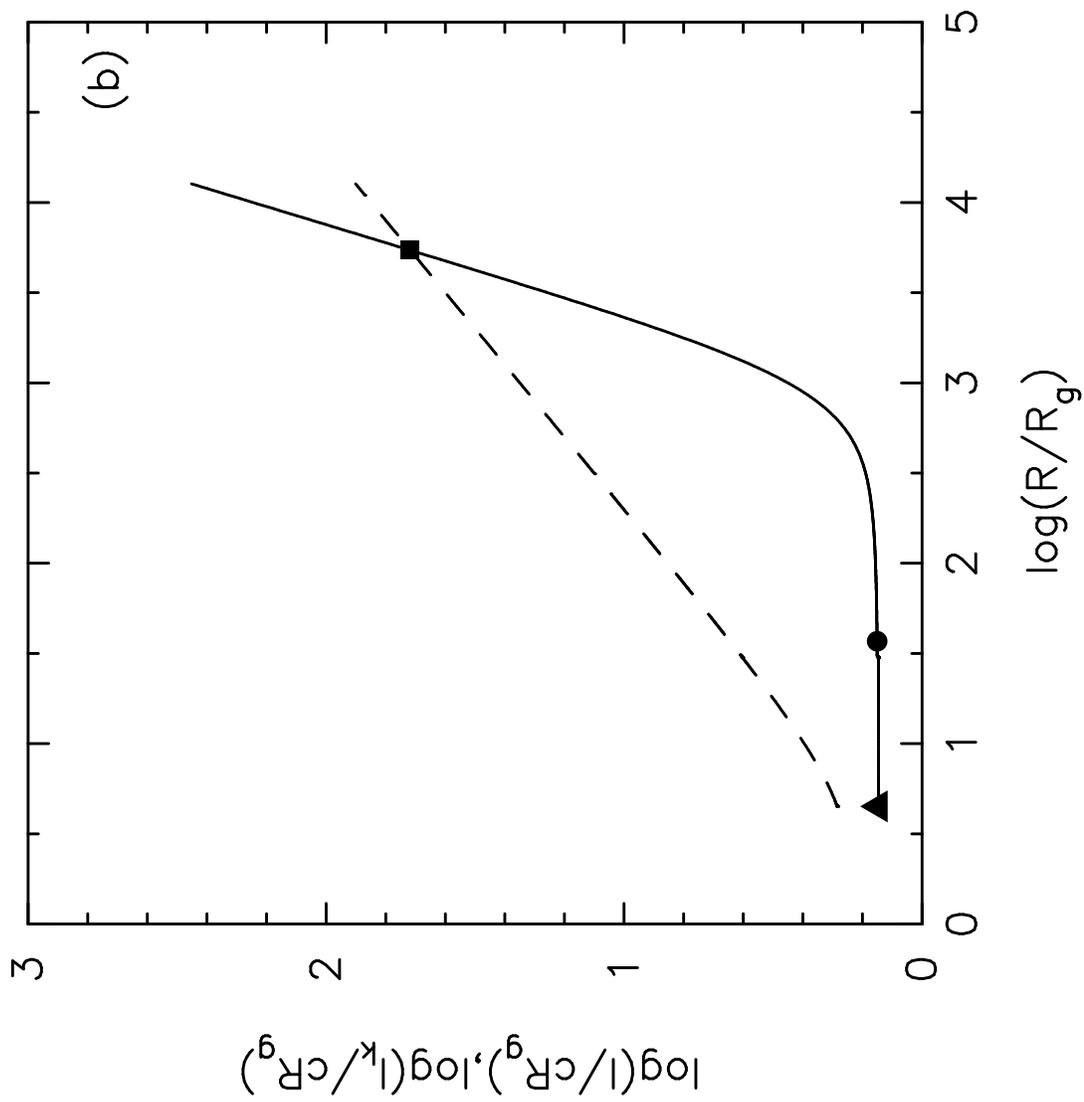
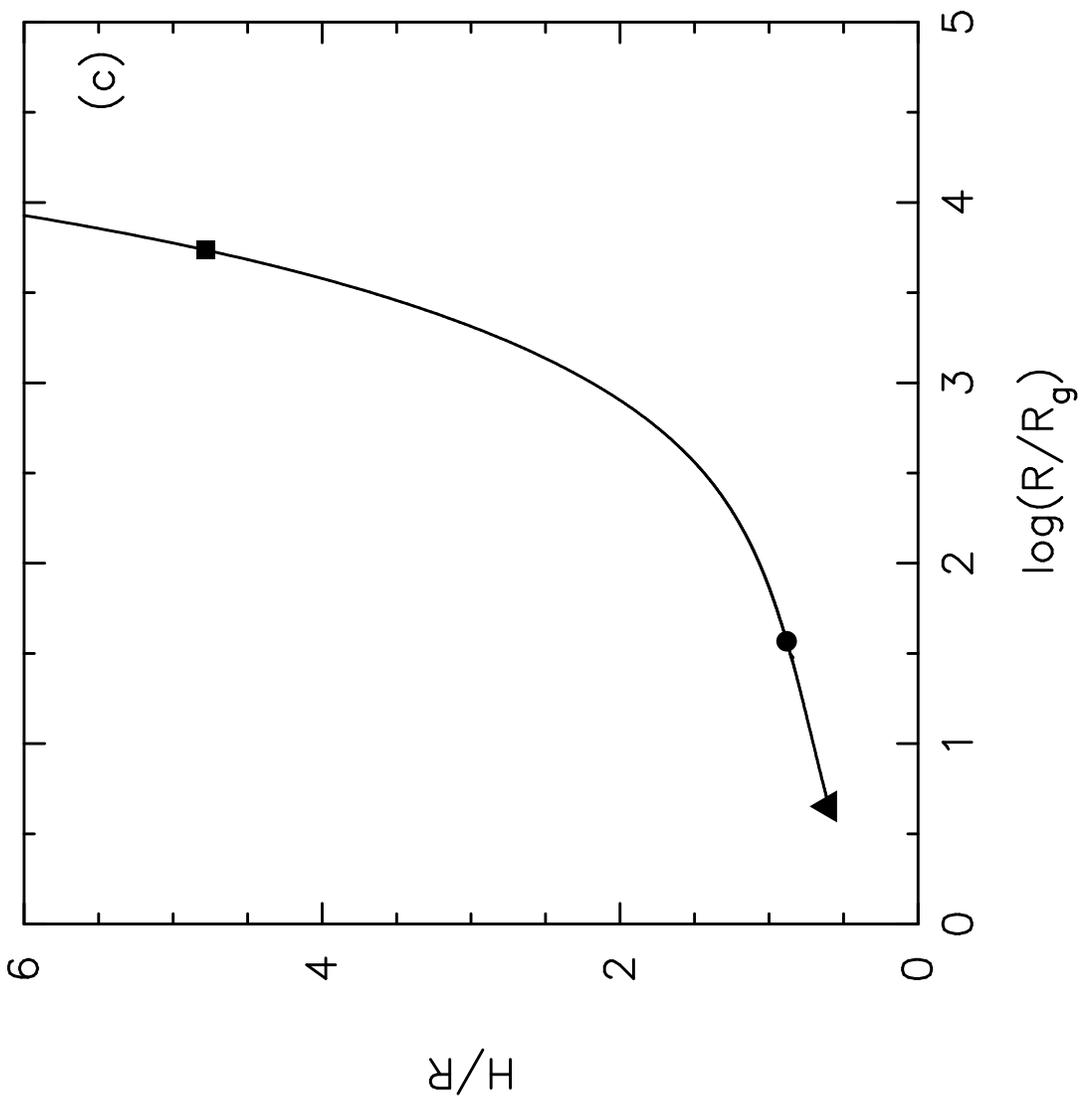
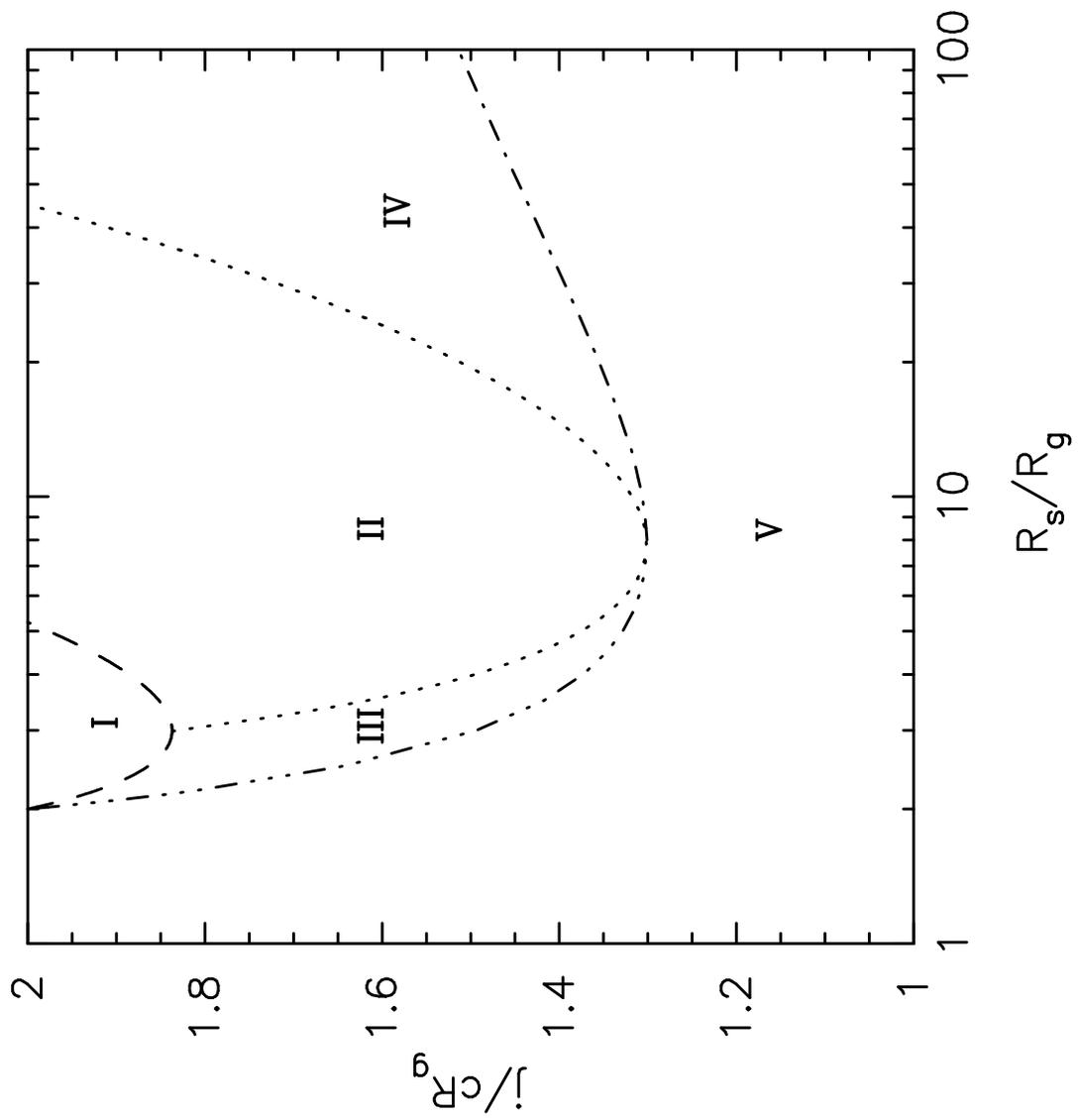
\end{document}